\documentclass[twocolumn,apl,showpacs]{revtex4}
\usepackage{subfigure}
\usepackage{color}
\usepackage{graphicx}

\begin{document}

\title{Discrete Coordinate Scattering Approach to Nano-optical Resonance}
\author{Jun Yang}
\affiliation{Department of Physics, Peking University, 100871, Beijing, P.R. China}
\author{H. Dong}
\email{dhui@itp.ac.cn} \affiliation{Institute of Theoretical
Physics, Chinese Academy of Sciences, 100190, Beijing, P.R. China}
\author{C.P. Sun}
\email{suncp@itp.ac.cn} \affiliation{Institute of Theoretical
Physics, Chinese Academy of Sciences, 100190, Beijing, P.R. China}

\begin{abstract}
In this letter, we investigate the coherent tunneling process of
photons between a defected circular resonator and a waveguide based
on the recently developed discrete coordinate scattering methods (L.
Zhou \textit{et al.}, Phys. Rev. Lett. \textbf{101}, 100501 (2008)).
We show the detailed microscopic mechanism of the tunneling and
present a simple model for defect coupling in the resonator. The
Finite-Difference Time-Domain(FDTD) numerical results is explored to
illustrate the analysis results.
\end{abstract}

\maketitle


Recently, resonant optical cavities has attracted lots of
interesting from both experimental and theoretical
aspect~\cite{vahla03review}. Devices based on these micro-cavities
has been widely used for processing optical
signals~\cite{joannopoulos}. In most hardware of these signal
processing, the elementary unit consists of a micro-cavity and a
side coupling waveguide~\cite{vahla03review}. This unit has been
experimentally realized based on photonic crystal, most recently
ring made with GeN nanowire~\cite{pdyang06prl}. With such kind
photonic setups, photon blockade was experimentally observed in this
unit~\cite{kimble08}. For this unit, it is well known that the
incident wave is totally reflected by the ring resonator when it is
resonant with the modes of the ring. Thus, the resonator takes the
role of frequency-selection~\cite{little98opt} in this unit.

Such transfer process in the unit is crucial to design more
complicated devices. To investigate this phenomena, one way is to
directly solve the Maxwell equations. However, it is impossible to
get an analytic result to reveal the key physical mechanism for even
a simple configuration. Thus, this is usually done by numerical
simulation, such as Finite-Differential-Time-Domain(FDTD)
method~\cite{taflove}, which is used as a validation approach.
Theoretically, coupled mode theory(CMT) is introduced to study the
resonator scattering properties in this
unit~\cite{little98lightware,little98opt}. However, this is done in
the mode space, and therefore not intuitive in real space.

In this letter, we generalize the most recently developed discrete
coordinate scattering methods~\cite{lan08} to investigate the
detailed scattering properties for the optical signal processing.
Here we model the resonator and the waveguide as array of coupled
cavities. And the coupling between resonator and waveguide is
modeled as hopping of the photon among cavities. We present the
method by a qualitatively analysis of an example of a notched-ring
resonator. Rigorous FDTD numerical simulations is also performed to
verify the obtained analytical result.

\begin{figure}[tbp]
\includegraphics[bb=54bp 584bp 461bp 774bp, width=8cm]{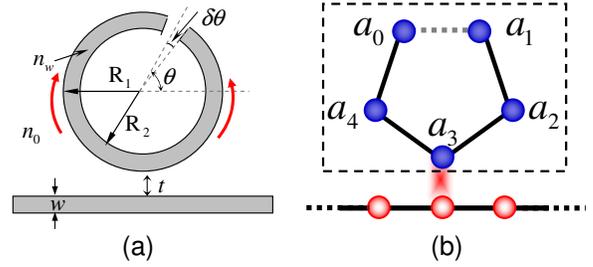}
\caption{(Color online) Frequency-selective filter. (a) A notched
ring is evanescently coupled to the waveguide. The notch couples the
CW and CCW traveling modes; (b) Model of the unit by discrete
coupled resonator array. The notch is characterized as point
inhomogeneous of the tunneling strength along the ring.}
\label{fig:resonator}
\end{figure}

\begin{figure}[tbp]
\includegraphics[width=6cm]{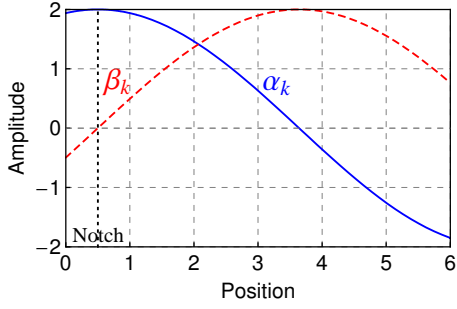}
\caption{Ez-field amplitude of $\alpha_k$(Blue line) and
$\beta_k$(Red dashed line) mode on the notched ring. The position of
notch is marked with dot line. For $\alpha_k$ mode, the amplitude of
Ez-field at the notch is at maximum. However, the amplitude is null
for $\beta_k$ mode.} \label{fig:modes}
\end{figure}

The elementary units under investigation is schematically
illustrated in Fig. \ref{fig:resonator}(a). The defected ring
resonator is evanescently coupled to the signal waveguide. For
perfect ring, it supports the clockwise(CW) and counter
clockwise(CCW) traveling wave and there is no interaction between
the two traveling wave modes. However, the scattering of the
defected notch causes the recombination of the CW and CCW modes. In
the context of CMT~\cite{little98opt}, the notched ring is mapped to
the coupled double-ring resonator and the notch in single ring plays
the role of the gap, which couples the double rings. Here, we treat
the defected circular resonator as the discrete coupled cavity array
with coupling constant different at the defected area, which is
illustrated in Fig.~\ref{fig:resonator}(b). The array consists of
$N$ resonators with tight-binding coupling. The notch in the ring is
marked with a dashed line. The Hamiltonian here reads
\begin{equation}
H_{c}=\sum_{j}\omega a_{j}^{\dagger }a_{j}-J\left( a_{j}^{\dagger
}a_{j+1}+\text{h.c.}\right) +\delta J\left( a_{0}^{\dagger
}a_{1}+\text{h.c.}\right) ,
\end{equation}%
where $a_{j}$ is the creation operator for photon with energy
$\omega $ the $j-th$ cavity and $J$ is the tunneling strength
between nearby cavities with defected induced inhomogeneous coupling
$\delta J$. By the Fourier transformation $a_{j}^{\dagger }=\sum
\exp \left(-ijk\right) c_{k}^{\dagger }/\sqrt{N}$, the Hamiltonian
in the momentum space reads $H_{c}=\sum_{k}\left( \omega -2J\cos
k\right) c_{k}^{\dagger }c_{k}+\delta J\sum_{k,p}\left[ \exp \left(
ip\right) c_{k}^{\dagger }c_{p}+h.c.\right] /N $. In the rotation
wave approximation, we keep only $p=\pm k$ term and obtain the
Hamiltonian as $H_{c}=\sum_{k>0}\varepsilon _{k}\left(
c_{k}^{\dagger }c_{k}+c_{-k}^{\dagger }c_{-k}\right)
+\sum_{k>0}2\delta J\left( \exp \left( -ik\right) c_{k}^{\dagger
}c_{-k}+\text{h.c.}\right) /N$, where $\varepsilon _{k}=\omega
-2\left( J-\delta J/N\right) \cos k$. Physically, $c_{k}$and
$c_{-k}$ stand for the CW and CCW traveling wave, which has been
introduced phenomenally in CMT. The coupling between the two modes
is induced by the defect of the inhomogeneous of tunneling $\delta
J$. Diagonalizing the Hamiltonian with unitary transformation
$c_{k}^{\dagger}=\exp \left(ik/2\right) (\alpha _{k}^{\dagger}+\beta
_{k}^{\dagger})\sqrt{2}$ and $c_{-k}^{\dagger}=\exp \left(-
ik/2\right) (\alpha _{k}^{\dagger}-\beta _{k}^{\dagger})\sqrt{2}$
for $k>0$, we obtain
\begin{equation}
H_{c}=\sum_{k>0}\varepsilon _{k}^{+}\alpha _{k}^{\dagger }\alpha
_{k}+\varepsilon _{k}^{-}\beta _{k}^{\dagger }\beta _{k},
\end{equation}%
where $\varepsilon _{k}^{\pm }=\left( \varepsilon _{k}\pm 2\delta
J/N\right) $ and
\begin{eqnarray}
\alpha_k^{\dagger} &=& \frac{1}{\sqrt{2N}} \sum_j 2a_j^{\dagger} \cos \left[(j-1/2)k\right] \\
\beta_k^{\dagger} &=& \frac{1}{\sqrt{2N}} \sum_j 2ia_j^{\dagger}
\sin \left[(j-1/2)k\right].
\end{eqnarray} are the eigen-modes of the notched ring.
And they were related to photonic molecule in
Ref.~\cite{bayer98prl,pdyang06prl}. Actually, this single defected
notch induced the split of the degenerate resonant peak of CW and
CCW traveling wave. The splitting can be probed by checking the
transmission rate in the waveguide, which will be investigated in
the following parts. Actually, those modes correspond to a symmetric
and antisymmetric state with respect to the notch. The amplitude for
the modes $\alpha_k$ and $\beta_k$ is illustrated in
Fig.~\ref{fig:modes}. In this discrete model of the ring resonator,
the position $j_{\mathrm{notch}}$ of notch should between the
resonator $a_0$ and $a_1$, namely $j_{\mathrm{notch}}=1/2$. Thus for
the $\alpha$ mode, the amplitude of the field at the notch is at
maximum. However, there is no field in the notch for the $\beta_k$
mode.

\begin{figure}[tbp]
\subfigure[]{\includegraphics[clip,width=6cm]{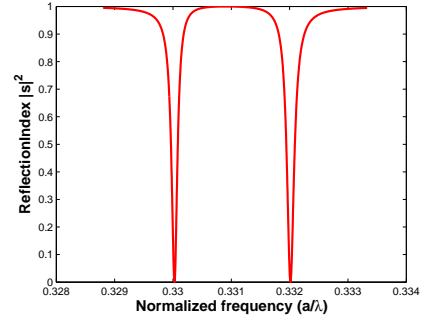}}
\subfigure[]{\includegraphics[width=6cm]{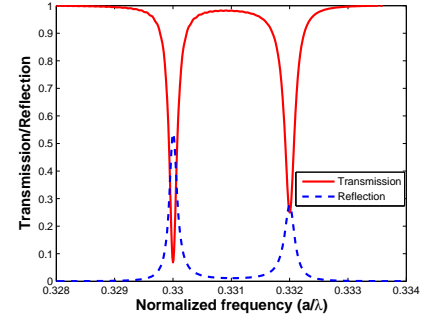}} \caption{Plot
of the reflection rate with the parameters
$\omega=\omega_1=8.772\times 10^{14}Hz$, $J=J'=3.3\times 10^{14}Hz$,
$g=3\times 10^{13}Hz$ and $\delta J=2.98\times 10^{13}Hz$. (a)
Theoretical calculation of the reflection of the filter; (b) FDTD
simulation of the reflection and transition rate. Two resonator
peaks appear at the cavity mode $\varepsilon_{k}^{\pm}$.}
\label{fig:reflec}
\end{figure}

To probe the optical property of the resonator, a transmission line
waveguide is introduced to couple the signal in the microcavity.
Physically, the waveguide can be optical fiber, nanowire and
defected lines on photonic crystal for different systems. Here, we
also model the transmission line waveguide as the discrete coupled
resonator, which is illustrated in Fig. \ref{fig:resonator}(b). Each
resonator couples to the near-by resonators by means of photon
hopping. Those coupled-resonator array has been proposed to be a new
type of optical waveguide\cite{yariv99opl}. The Hamiltonian for the
transmission line waveguide here reads
\begin{equation}
H_{l}=-J^{\prime}\sum\left(b_{j}^{\dagger}b_{j+1}+\text{h.c.}\right)+\omega^{\prime}b_{j}^{\dagger}b_{j},
\end{equation}
where $b_{j}^{\dagger}$ is the bosonic creation operator for photon
on the $j-th$ resonator. The tight-binding Hamiltonian here is a
good approximation for modeling the coupled resonator array, which
is proved in Ref. \cite{yariv99opl}. And in Ref. \cite{lan08}, it is
showed the model is equivalent to the simple dielectric waveguide
under the linear dispersion area with right going wave $\phi_R(x)$
and left going wave $\phi_L(x)$ as showed in Ref. \cite{fan98}. The
waveguide couples to the cavity at the $0$-th resonator as the
tunneling between 0-th resonator and the $j_{0}$-th resonator of the
ring resonator
$H_{I}=g\left(a_{j_{0}}^{\dagger}b_{0}+a_{j_{0}}b_{0}^{\dagger}\right)$.
The total Hamiltonian for this elementary unit reads
$H=H_{c}+H_{l}+H_{I}$. For the single photon case, the eigenwave
stationary scattering function can be written as
\begin{equation}
\left|\Phi\right\rangle
=\sum_{j}u_{j}b_{j}^{\dagger}\left|0\right\rangle
+\sum_{k>0}\left(v_{k}\alpha_{k}^{\dagger}+w_{k}\beta_{k}^{\dagger}\right)\left|0\right\rangle
,  \label{eq:wavefunc}
\end{equation}
where $u_{j}$ is the amplitude of photon in the $j-th$ resonator on
the waveguide and $v_{j}$($w_{j}$) is the amplitude of photon on the
resonator with mode $\alpha^{\dagger}$($\beta^{\dagger}$).
Substituting the stationary scattering function into the
time-independent Schr\"{o}dinger equation $H\left|\Phi\right\rangle
=E\left|\Phi\right\rangle $, one can obtain equations for the
amplitude of the site as
\begin{widetext}
\begin{eqnarray}
-J^{\prime }\left(u_{j+1}+u_{j-1}\right)+\omega^{\prime
}u_{j}+\frac{
g\delta_{j,0}}{\sqrt{2N}}\sum_{k>0}\left(v_{k}+w_{k}\right)e^{-i\left(1/2-j_{0}\right)k}+\left(v_{k}-w_{k}\right)e^{-ik(j_{0}-1/2)}
& = & Eu_{j}, \label{eq:wavesub1}\\
\varepsilon_{k}^{+}v_{k}+\frac{gu_{0}}{\sqrt{2N}}\left(e^{i\left(1/2-j_{0}\right)k}+e^{i(j_{0}-1/2)k}\right) & = & Ev_{k}, \label{eq:wavesub2}\\
\varepsilon_{k}^{-}w_{k}+\frac{gu_{0}}{\sqrt{2N}}\left(e^{i\left(1/2-j_{0}\right)k}-e^{i(j_{0}-1/2)k}\right)
& = & Ew_{k}.\label{eq:wavesub3}
\end{eqnarray}
\end{widetext}

When the incident energy coincides with the energy of $\alpha$ mode,
namely $E = \varepsilon_k^+$, the amplitude $u_0=0$ by
Eq.~(\ref{eq:wavesub2}) and $w_k=0$ by Eq.~(\ref{eq:wavesub3}). Thus
only $\alpha$ mode is excited here and the Ez-field at the notch is
at maximum. When the incident energy is resonant with the energy of
$\beta$ mode, namely $E = \varepsilon_k^-$, the amplitude $u_0=0$ by
Equation(\ref{eq:wavesub3}) and $v_k=0$ by
Equation(\ref{eq:wavesub2}). Here, only $\beta$ mode is excited and
the amplitude of the field is zero. This phenomena has been
numerically predicted in Ref.~\cite{little98opt}.

For the scattering process, the photon incidents from the left side
of the line waveguide and is partially reflected by the notched ring
resonator array. The amplitude in the waveguide reads $u_{j}=\left[
\exp \left( iqj\right) +r\exp \left( -iqj\right) \right] \theta
\left( -j\right) +s\exp \left( iqj\right) \theta \left( j\right) $ ,
where $r$($s$) is the reflection(transmission) coefficient of the
photon. The coefficient of the transition here is obtained
\begin{equation}
s=\left( 1-\frac{Q}{2iJ^{\prime }\sin q}\right) ^{-1},
\end{equation}%
where $Q=2g^{2}/N\sum_{k>0}\cos ^{2}(k(j_{0}-1/2))/\left(
E-\varepsilon _{k}^{+}\right) +\sin ^{2}(k(j_{0}-1/2))/\left(
E-\varepsilon _{k}^{-}\right) $. The reflection rate $1-|s|^{2}$
peaks at the resonance with the internal ring mode $\varepsilon
_{k}^{\pm }$, where the optical signal is totally reflected. To see
this resonant filtering effect, we plot the transmission rate
$|s|^{2}$ verse the the incident energy of photon in Fig.
\ref{fig:reflec}(a). For comparison, a rigorous FDTD~\cite{taflove}
simulation is performed with free available code F2P~\cite{qiu} for
Transverse-Magnetic mode. The simulation is performed with the
following parameter: $n_0=1$, $n_w=3.2$, $R_1=5a$, $R_2=4.4 a$,
$\theta=54^{\mathrm{o}}$ and $\delta\theta=1.5^{\mathrm{o}}$, where
$a=0.25 \mu m$.  The transmission and reflection rate is plotted in
Fig.~\ref{fig:reflec}(b) around the resonant frequency $\varepsilon
_{k}=0.331 \times 2 \pi c/a $. In Fig.~\ref{fig:reflec}(a), the
parameters for this defect ring resonator unit are chosen to fit the
numerical simulation: $\omega=\omega_1=8.772\times 10^{14}Hz$,
$J=J'=3.3\times 10^{14}Hz$, $g=3\times 10^{13}Hz$ and $\delta
J=2.98\times 10^{13}Hz$. Therefore, our model confirms the function
of the filter and its mechanism. However, we should notice a
drawback to this approach: the quantitative analysis will relies on
the numerical simulation. At present, the coupling constant at the
notch and between the transmission line and the resonator should be
fitted by numeric simulation.

In conclusion, we have presented a theoretical investigation of the
notched ring filter with discrete coordinate scattering methods. The
detailed mechanism of the coupling between the clockwise and
count-clockwise is revealed to be the inhomogeneous tunneling in the
resonator array system. We have predicted some experimentally
accessible results for the coherent transmission and reflection
property and present detailed explanation to field amplitude
distribution at the notch.

This work is supported by NSFC No.10474104, No. 60433050, and No. 10704023,
NFRPCNo. 2006CB921205 and 2005CB724508.

\end{document}